\documentclass[aps,prd,twocolumn,groupedaddress,showpacs]{revtex4}
\usepackage{graphics}

\begin{document}


\title{Exact Dragging of Inertial Axes 
       by Cosmic Energy-Currents 
       on the Past Light-Cone: Mach's Principle}

\author{Christoph Schmid}
\email{chschmid@itp.phys.ethz.ch}

\affiliation{ETH Zurich, Institute for Theoretical Physics, 
8093 Zurich, Switzerland}

\date{\today}

\begin{abstract}
We prove exact rotational dragging
of local inertial axes ($\equiv$ spin axes of gyroscopes) 
by arbitrary cosmic energy-currents 
on the past light-cone of the gyroscope
for linear perturbations of Friedmann-Robertson-Walker cosmologies.
Hence, the principle
formulated by Mach holds for arbitrary
linear cosmological perturbations.

\end{abstract}


\maketitle


\section{The hypothesis formulated by Ernst Mach}

Experimentally spin axes of gyroscopes
           directly give 
           the time-evolution of local inertial axes
           (as in inertial guidance systems).
Conversely,    relative to local inertial axes 
           there is  
           no gyroscope-precession.
This is a {\it local} fact.

In a {\it cosmological} context, we have a 
super-precise {\it observational fact:}
Spin-axes of gyroscopes do not precess 
relative to quasars,
except for an extremely small dragging effect by Earth-rotation,
the Lense-Thirring effect, 
which is negligible for gyroscopes away at a few Earth radii.

The question addressed in Mach's principle:
{\it What physical cause} determines 
the time-evolution for spin-axes of gyroscopes,
i.e. the time-evolution of inertial axes?
In the words of John A. Wheeler: 
 ``Who gives the marching orders'' 
to gyroscope axes 
($\equiv$ inertial axes)?

The {\it  postulate formulated by Mach} \cite{Mach}:
Inertial axes  exactly follow 
          an average of the motion of cosmological masses:
          {\it exact frame-dragging}.

Since Newton's gravitational force
cannot exert a torque on a gyroscope,
Mach wrote: it is unknown, 
            what new force
            could do the job.

In General Relativity, 
  gravito-magnetism  
causes the Lense-Thirring effect,
{\it extremely small torques} on  gyroscopes in orbit arond the Earth
(caused by the Earth's rotation) detected by Gravity Probe B.
In striking contrast, {\it Mach postulated exact dragging},  
not a little bit of dragging.

Mach wrote that he did not know, {\it what average} of
            cosmological masses and their motions should be taken.
I. Ciufolini and J.A. Wheeler wrote in 1995 \cite{Ciufolini.Wheeler}
that it is {\it still unknown} what average should be taken.

\section{Results}
%
     In Refs.~\cite{Rovaniemi, CS.1, CS.2}
we have proved
exact dragging of inertial axes  
       by cosmic energy-currents (Mach's principle)       
on {\it space-like slices} 
(slices connecting points of equal local Hubble expansion-rate)
for all possible linear perturbations
of all Friedmann-Robertson-Walker (FRW) backgrounds.

Our new results: We prove
exact dragging of local inertial axes
at any space-time point $P_0$
by   cosmic energy-currents 
on the {\it past light-cone}
of the gyroscope-observation at $P_0$ 
for all possible linear perturbations of spatially flat FRW
backgrounds.  
  
The angular momentum constraint at $P_0$   
             from the
             {\it past light-cone} of $P_0$ for linear perturbations
             gives a linear {\it ordinary differential equation}
             in the radial variable.

The solution of the {\it angular momentum constraint} from the 
past light-cone gives {\it nothing more} and {\it nothing less}
than (1) the proof of {\it exact dragging} of 
inertial axes by cosmic energy currents,
(2) the form of the {\it dragging weight-functions} 
for various Hubble-rate histories.
%

\section{Past light-cone coordinates for an 
unperturbed FRW universe}
%

In an unperturbed and spatially flat
Friedmann-Robertson-Walker (FRW) universe, 
we single out one comoving observer, 
and we choose the spatial origin at his position.
The comoving distance $\chi$ and the conformal time $\eta$ are defined
(with $c \equiv 1$),
\begin{eqnarray}
\chi &\equiv& {\mbox {comov. distance from observ.}} 
\, \, \, \, \, \,  
\chi \equiv \, \, r/a(t),
\nonumber
\\
\eta &\equiv& {\mbox{conformal time,}} \quad \quad \quad  \quad \quad \quad
\, \, \, 
d\eta \equiv dt/a(t),
\label{conformal.time}
\end{eqnarray}
where 
$r$ is the measured radial distance from the origin at fixed Hubble-time,
$dt$ is the measured time interval in a comoving frame, and 
$a(t)$ is the scale factor.
The scale factor is set to one 
at $t_0 \equiv$ observation-time, $a(t_0) \equiv 1.$
The light cones are at 45 degrees in the $(\eta, \chi)$-plane,
therefore $(\eta, \chi)$ is a conformal pair.

In the retarded Green function
for a given conformal observation-time $\eta$  
at the spatial origin,
the earlier 
conformal source-emission-time       $\eta\,'$ at 
comoving  source-emission-distance   $\chi\,'$  
on the past light-cone of the observation is
\begin{eqnarray}     
\eta\,' &=& (\eta - \chi\,') \quad \quad \quad {\mbox{with}} \quad  c \equiv 1.
\nonumber
\end{eqnarray}
%
The conformal {\it past-light-cone coordinate} $v$
is defined,
%
\begin{eqnarray}
v &\equiv& \, \eta + \chi. 
\end{eqnarray}
The coordinate $v$ is 
constant 
on each of the past-light cones of the chosen observer,
i.e. $v$ labels the past light-cones for our chosen observer.
At the position of the observer, $v$ is equal to the conformal time $\eta$.
In the integration over sources for {\it retarded potentials},
$v$ is  fixed, and $(\chi. \theta, \phi)$ are the integration variables.

For an unperturbed and spatially flat 
FRW universe, 
the metric in  past-light-cone coordinates $(v, \chi, \theta, \phi)$ is
\begin{eqnarray}
ds^2 &=&   
a^2(\eta)_{\eta = v - \chi} \, 
[- dv^2 + 2 \, dv \, d\chi + \chi^2 \,  d \omega^2],
\nonumber
\end{eqnarray}
where $d\omega^2$ is the line element on the unit 2-sphere,
\begin{eqnarray}
d \omega^2 &\equiv& d \theta^2 + \sin^2 \theta \, d \phi^2.
\nonumber
\end{eqnarray}
The non-zero components of the unperturbed metric are
\begin{eqnarray}
 g_{vv}^{(0)}      = - a^2,          \, \, \, \,
&&  \, \,   g_{v \chi}^{(0)} = a^2,
\nonumber
\\ 
 g_{\theta \theta}^{(0)} = (a \chi)^2,  \, \, 
&&  \, \,   g_{\phi \phi}^{(0)} = (a \chi \sin
 \theta)^2,
\nonumber
\\ 
(- {\rm det} g_{(0)})^{1/2}  \, 
\equiv &(- g_{(0)})^{1/2}&   =  
 \,   a^4 \chi^2 \sin \theta.
\label{metric.unperturbed.final}  
\end{eqnarray}
Note that  $g_{\chi \chi}^{(0)} = 0,$   
because along a world line of a photon $ds^2 = 0,$
and for a photon observed at $\chi = 0$, 
$(v, \theta, \phi)$ is fixed, while $d \chi \neq 0.$     

The inverse of the unperturbed metric
is non-trivial only 
for the inverse of the (2x2)-matrix in the $(v, \chi)$-sector.
The nonzero elements of the unperturbed inverse  metric are
\begin{eqnarray}
g^{\chi \chi}_{(0)} = \frac{1}{a^2},  
&& \quad g^{v \chi}_{(0)} =  \frac{1}{a^2},  
\nonumber
\\ 
g^{\theta \theta}_{(0)} =  \frac{1}{(a \chi)^2},  
&& \quad 
g^{\phi \phi}_{(0)}    =  \frac{1}{(a \chi \sin \theta)^2}.
\label{unpert.inv.metric.FRW.x} 
\end{eqnarray}
Note that $g^{vv}_{(0)} = 0.$

\boldmath
\section{Vector spherical harmonics $\vec{X}^{\pm}_{\ell m}$}
\unboldmath

Vector spherical harmonics
form a basis for vector fields tangent to 2-spheres.
They
have been discussed in detail in Section~IV of 
    \cite{CS.2}.

The vector spherical harmonics of {\it Regge and Wheeler}
$\tilde{x}^{\pm}_{\ell m}$
   \cite{Regge.Wheeler}
are defined by
%
\begin{eqnarray}
\tilde{x}^+_{\ell m}  
\equiv  d \, Y_{\ell m} \quad 
&\Leftrightarrow& \quad (x^+_{\ell m})_{\alpha} 
\equiv \partial_{\alpha} Y_{\ell m},
\\ 
\tilde{x}^-_{\ell m}  
\equiv - ^{(2)}*   d \, Y_{\ell m} \quad 
&\Leftrightarrow& \quad (x^-_{\ell m})_{\alpha} 
\equiv - \varepsilon_{\alpha \beta} g^{\beta \gamma}   \partial_{\gamma}
Y_{\ell m}.
\nonumber
\end{eqnarray}
%
On the left is the abstract notation of differential forms,
on the right is the explicit component notation:
$\tilde{x}^+_{\ell m}$ is the gradient of $Y_{\ell m},$ while
$\tilde{x}^-_{\ell m}$ is its Hodge dual on the 2-spheres, denoted by
$^{(2)}*$. 
On the 2-sphere of any radius, 
the Levi-Civita tensor is  $\varepsilon_{\alpha \beta}$.
The vector spherical harmonics of Regge and Wheeler
have covariant components ($\equiv$ 1-form components)
{\it independent of the radial coordinate} $\chi$, 
%
\begin{equation}
\partial_{\chi} (x^{\pm}_{\ell m})_{\alpha} = 0,  
\quad \quad \alpha = (\theta, \phi).
\end{equation}
%

In contrast,
the {\it physical vector spherical harmonics} $\vec{X}^{\pm}_{\ell m}$
used in classical electrodynamics 
    \cite
{Jackson}
have a {\it point-wise norm} 
$\, g(\vec{X}^{\pm *}_{\ell m}, \vec{X}^{\pm}_{\ell m})$
{\it independent of the radial coordinate} $\chi$,
and therefore they have 
{\it LONB-components} (denoted by hats) 
{\it independent of the radial coordinate},
%
\begin{eqnarray} 
\nabla_{\chi} \vec{X}^{\pm}_{\ell m} &=& 0,  
\quad \quad \partial_{\chi} \, (X^{\pm}_{\ell m})_{\hat{k}} = 0.
\end{eqnarray}
%

The parity is $P = (-1)^{\ell}$ for $\vec{X}^+_{\ell m}$ (``natural parity''),
while  $P = (-1)^{\ell +1}$ for $\vec{X}^-_{\ell m}$ (``unnatural parity'').

As shown in the next section, 
the precession of a gyroscope at the origin can be caused 
only by cosmological energy currents
with $\ell~=~1$ and 
parity  $P=+1$,
i.e. with a superscript minus for the
unnatural parity sequence.
For a given  source-radius $r_{\rm s},$ in the Green function, 
we can specialize to 
$m~=~0$  
without loss of generality. 

For $m = 0,$ 
one has rotational symmetry around the $z$-axis,
and the vector field $\vec{X}^{-}_{\ell, m = 0} $
points in the $\phi$-direction.
If the vector field is a 3-velocity field,
the LONB component $V_{\hat{\phi}}$ is the measured 3-velocity in the 
$\phi$-direction, and the contravariant component $V^{\phi}$
is the measured angular velocity around the $z$-axis, $d \phi/dt$.

For $\ell =1$ (with $m = 0, \, P = +1$),
the angular velocity around the $z$-axis is independent of $\theta$.
This is a {\it rigid rotation} arond the $z$-axis 
with angular velocity  $\Omega = (d \phi /dt) = v^{\phi}$.
Using 
$Y_{\ell =1, m=0} = \sqrt{3/(4 \pi)} \, \cos \theta,$   
%
\begin{eqnarray}
\vec{X}_{\ell = 1, m = 0} &\equiv& \sqrt{3/(4 \pi)}  \, \, \vec{V},
\nonumber
\\
V^{\phi}  \, &=& \,  1,  \quad \quad \quad \, \, {\mbox{angular velocity}},
\nonumber
\\ 
V_{\hat{\phi}} \,  \, \, &=&  \,  \sin \theta,  \quad \quad {\mbox{velocity}}.
\end{eqnarray}
%

For any $(\ell, m)$, the unnatural-parity vector spherical-harmonics 
$X^-_{\ell, m}$ are called {\it toroidal}.\\

\boldmath
\section{Precession of gyroscope-spin caused by toroidal
vorticity perturbations with $\ell = 1$}
\label{gyroscope.precession.toroidal.vorticity}
\unboldmath

We treat all
linear perturbation fields 
on a spatially flat
FRW background ($K = 0)$, 
and
    {\it all}
energy-momentum-stress tensors, 
i.e. 
all types of matter, not necessarily of the perfect-fluid form, 
dark energy, and a cosmological constant,
and   
    {\it all} 
field configurations
of observed energy currents 
$J^{\varepsilon}_{\hat{k}} \equiv T^{\hat{0}}_{\, \, \hat{k}}.$
We followed this general approach already in our papers
 \cite{Rovaniemi, CS.1, CS.2},  
which is in striking contrast 
to the other literature,  
which only treated 
the artificial situation of
spherical shells of matter rotating rigidly around one given axis.

Linear cosmological perturbations 
can be decomposed into 
{\it scalar,} {\it vector,} and {\it tensor} sectors
as discussed by Bardeen in 1980 
\cite{Bardeen.1980}:

(1) In the {\it scalar sector}, 
3-vector fields are gradients of scalar fields,
the curl is zero, the fields are determined by their divergence.
Traceless symmetric 3-tensors of second rank 
are obtained from scalar fields 
by 3-covariant derivatives.  

(2) In the {\it vector sector}, 3-vector fields are divergenceless,
given by the curl, i.e.  {\it vorticity}.
Therefore the vector sector 
is also called vorticty sector.
Symmetric 3-tensor fields of second rank 
are obtained from vorticity vector fields
by 3-covariant derivatives, and they are traceless but not divergenceless.

(3) In the {\it tensor sector}, 
traceless, divergenceless 3-tensor fields
describe gravitational waves.

There is an important difference between our problem,
the {\it angular momentum constraint}, and 
Bardeen's problem 
   \cite{Bardeen.1980}: 
In our problem,
the {\it position of the gyroscope} at $P_0$ is singled out.
Relative to the gyroscope, 
the decomposition in eigenstates of 
{\it angular momentum} and {\it parity} 
and correspondingly 
the decomposition of the vector sector ($\equiv$ vorticity sector)
in {\it toroidal vorticity} versus {\it poloidal vorticity} 
is  extremely useful: 

(2a) {\it Toroidal vorticity} fields 
are defined to have {\it unnatural parity,} $P = (-1)^{\ell + 1}.$
Only this sector causes the {\it precession of gyroscopes}
and {\it rotational dragging}. 
The simplest example  
of a toroidal vorticity field 
is the velocity field of a rotating shell of matter,
which has $(\ell =1, P = +1).$

(2b) {\it Poloidal vorticity} fields
have {\it natural parity,}  $P = (-1)^{\ell}.$
The simplest example of a poloidal vorticity field 
is the electric current in the wire wound around an iron ring.


We shall show that for our problem, the dragging of gyroscope's axes 
by cosmic energy currents,
the mathematics 
of totally general linear perturbations  fields
is equivalent to the mathematics of 
the special case
of spherical shells of matter at every radius
around our selected observer with his gyroscopes, 
with every shell in rigid rotation around a different rotation axis.
This is shown using   three theorems, 
which are based on the {\it symmetries} 
relevant for Mach's principle, {\it rotation} and {\it parity,}
%
\begin{enumerate}
\item  The 
         precession 
       of a 
         gyroscope 
       (relative to the local axes chosen by a given observer)
         {\it cannot} 
       be caused by 
         {\it scalar}
       perturbations nor,
       in linear perturbation theory,  
       by 
        {\it tensor} 
       perturbations,
       because the energy-currents of gravitational waves 
       are of second order in the gravitational field. 
\item
       In the {\it vorticity sector} ($\equiv$ vector sector),
       the precession of a gyroscope can be caused only 
       by energy-current fields $\vec{J}_{\varepsilon}$
        with $J^P = 1^+$ 
       relative to the given gyroscope's position,
       i.e. by 
         {\it toroidal}
       vorticity and with $\ell =1.$
\item  On every mathematical spherical shell 
       centered on the gyroscope considered:
       The energy-current field-component  
       which is toroidal and has $\ell = 1$
       (relative to the gyroscope) 
       is given by an  
           {\it equivalent rigid rotation}  
       with an 
           {\it equivalent angular velocity} of matter 
       $\vec{\Omega}^{\, \rm matter}_{\, \rm equiv} (\chi_{\rm s}).$
       The equivalent angular velocity of matter 
       is given by the 
            global inner product (scalar product)
       of  the energy-current field $\vec{J}_{\varepsilon}$ 
       with the toroidal fields 
       $\vec{X}^{-}_{\ell =1,m}$ on the shell of radius~$\chi,$
%
\begin{eqnarray}
&& \int d\Omega <\vec{X}^{- \, *}_{\ell =1,m} \, , \,
\vec{J}_{\varepsilon} (\chi, \theta, \phi)>
\, \, \equiv  \, (J_{\varepsilon})^-_{\ell=1,m}(\chi)
\nonumber
\\ 
&& = - \, \sqrt{16 \pi/3} \, \, 
(\rho_0 + p_0) \, R(\chi) \, \,
[\Omega_m (\chi)]^{\rm matter}_{\rm equiv},
\label{projection}
\end{eqnarray}
where $<... \, , \, ...>$ denotes the point-wise inner product,
and $d\Omega$ is the element of solid angle,  
while $\Omega_m$ denotes spherical-basis components 
of the angular velocity. 
In the $m=0$ sector, 
the energy current $\vec{J}_{\varepsilon}$ is given by 
$J_{\varepsilon}^{\phi} = T^{t \phi} = a T^{\eta \phi} = a T^{v \phi}.$
%
\end{enumerate}
%

The proofs of theorems (1) and (2) use the following facts:
The precession of a gyroscope-spin $d \vec{S} / dt$ 
relative to given local axes, 
which equals the 
   torque 
on the gyroscope, 
is an 
     axial vector, 
$J^P = 1^+.$~--- 
     For
       {\it scalar perturbations}
     all fields are derived from scalar fields via differentiation,
     but this can only produce 
     source-fields $\vec{J}_{\varepsilon}$ 
     in the natural parity sequence, $0^+, 1^-, 2^+,$ etc,
     which cannot contribute to the precession.~---
     For
         {\it tensor perturbations}
     (gravitational waves),
     all linear perturbations are given by a
     traceless, divergenceless 3-tensor, 
     from which one cannot form an axial vector field
     at the origin.


The proof of theorem (3) uses the following facts: 
For a general energy-momentum-stress tensor 
(not necessarily of the perfect-fluid type),
the local center-of-mass 3-velocity $\vec{v}$ 
in the toroidal vorticity sector with $m = 0$
is given by 
$ v^{\phi} = (\rho_0 + p_0)^{-1} a T^{v \phi}$ 
in linear perturbation theory, where $\rho_0$ and 
$p_0$ refer to the unperturbed FRW background.
 
In the Green function,
for a source at given $\chi_{\rm s}$,
a toroidal velocity field with $\ell = 1$ 
is a flow equivalent to a rigid rotation of a shell of matter, and
we shall choose the $z$-axis to be along the 
this shell-rotation axis, 
hence $m = 0.$\\   
%

\section{Minkowski corridors along 
incoming world-lines of photons}
\label{sect.Mink.corridor}

In this section, 
the discussion is {\it exact (non-perturbative)} and
{\it without background geometry}.
But we shall describe the procedure 
in a general cosmolgical language 
(no FRW background assumed).

We shall {\it define  coordinates} $x^{\mu}_P$  for each {\it event} $P$ 
by {\it measured} (observed) quantities. 
Therefore, {\it no gauge ambiguities} can arise.

We first choose an observation event $P_0$
anywhere in space-time.
Next, we choose a {\it Local Orthonormal Basis} (LONB) at $P_0$
{\it fixed by measurements (observations):}
\begin{enumerate}
\item
Choose  an observer with his 4-velocity $\bar{u}$ 
defining
the time-like basis vector of the 
Local Ortho-Normal Basis (denoted by hats over indices),
\begin{eqnarray}  
\bar{e}_{\hat{0}}(P_0)  &\equiv& \bar{u}_{\, \rm observer} (P_0).
\nonumber
\end{eqnarray}
We choose
the {\it observer} to be  at {\it  rest} 
in the {\it asymptotic Hubble frame}, 
i.e. in the Hubble frame
of quasars within some chosen fiducial volume 
at large luminosity distances $d_1 < d < d_2$ and over all angles.
This  means that the motion of the observer is chosen such that 
his observed dipole moment of radial quasar-velocities vanishes.
This construction goes through, when the universe is far from a FRW universe. 
\item
The direction 
of arriving photons from a chosen {\it ``north-pole quasar''}
fixes the local basis vector $\bar{e}_{\hat z}(P_0)$.
\item
The direction of arriving photons from
a chosen {\it ``zero-longuitude quasar''} 
fixes the {\it celestial null-meridian} 
$\phi =0$ on the celestial sphere,
which in turn fixes the local basis vector $\bar{e}_{\hat x}(P_0)$.
\item 
The fourth basis-vector, $\bar{e}_{\hat{y}}$, must be Lorentz-orthogonal 
to the three other basis vectors.
\end{enumerate}
This completely fixes the four basis vectors of our 
LONB($P_0$) directly by cosmological observations.

We now fix the {\it coordinates} 
of each {\it event} on the past light-cone of $P_0$,
one radial and two angular coordinates,
i.e. we uniquely fix
the mapping
\begin{eqnarray}
{\mbox{event}} \, P  \, \, &\Rightarrow&  \, \, x^{\mu}(P), 
\quad \quad \quad \mu = (1, \, 2, \, 3)
\nonumber
\end{eqnarray}
{\it directly by measurements}.

The angular coordinates $(\theta, \phi)$ are constant (by construction)
along every photon-world-line (geodesic) incoming at $P_0$.
The {\it event-coordinates} $(\theta_P, \phi_P)$ are equal to the 
{\it arrival directions observed} at $P_0$ 
of photons emitted at the event $P$,
\begin{eqnarray}
&& {\mbox{emission-event coord.}} \, \, 
(\theta_P, \, \phi_P)_{\rm emission} 
\nonumber
\\
&& = \, {\mbox{observation angles}} \, \,   \, \,
(\theta, \, \, \phi)_{\rm observed}(P_0).
\nonumber
\end{eqnarray}
%

For assigning a radial coordinate to every event $P$, 
an extremely useful concept
is the {\it Minkowski corridor} with a choice of {\it Minkowski coordinates}
along the world-line of one photon.
We start from the local Minkowski coordinate system 
around the observation event $P_0$, 
which is 
valid including  first derivatives at $P_0$ 
of the metric $g_{\mu \nu}$.
These Minkowski coordinates can be {\it extended along any one line} 
in space-time (geodesic or non-geodesic)
as long as the line is not self-intersecting.

The book cover of Misner, Thorne, and Wheeler 
   \cite{MTW}
shows the analogous concept of Euclidean corridors 
on the surface of an apple, 
and the 
first few pages of that book
discuss this concept.

In our case, we extend the Minkowski coordinates of our observer at $P_0$
with his LONB
in a Minkowski corridor along the world-line of one photon arriving
at $P_0$ from an
emission event $P$,
and we denote the radial coordinate of the event $P$ by $r_P$,
\begin{eqnarray}
r_P &\equiv &{\mbox{radial distance of event in Minkowski coordinates}}   
\nonumber
\\ 
&&  {\mbox{of our chosen observer at}} \, \, P_0
\nonumber
\\
&& {\mbox{along Minkowski corridor of world-line of photon}}.
\nonumber
\end{eqnarray}
For the astronomer, $r_P$  is measured by the luminosity distance 
of an object, e.g. $r_P = 100$ Mpc.
In classical electrodynamics \cite{Jackson}, 
the spatial distance $r_P$ of an event on the past light-cone
is the integration variable in retarded potentials.~--- 
The {\it spatial separation} $(dr)_{PP'}$ must be distinguished
from the {\it Lorentz-invariant space-time separation} $(ds^2)_{PP'} $,
which is zero for events on a photon world-line.
The 4-distance between two events is Lorentz-invariant, 
i.e. it is the same for all observers.
In contrast the 3-distance depends on our chosen observer 
(resp.  the 
output-observer in retarded potentials). 
The choice of an observer (at rest relative to asymptotic quasars)
{\it induces} a {\it spatial metric} along any incoming photon world-line.

This completes the determination of the mapping from any event $P$ to
event-coordinates $(r_P, \, \theta_P, \, \phi_P)$ directly by measurements.

Our procedure in this section has been in the 
spirit of defining Riemann normal coordinates in 3-space,
which uses geodesics emerging from a point $P_0$.
Riemann normal corrdinates refer to
Riemannian space (purely spatial coordinates, no time), 
while our coordinates refer to the past light-cone.

The {\it apex point} of the past light-cone creates no difficulties 
in the retarded potentials of classical electromagnetism,
and it creates no difficulties in our constraint equations.

The past light-cone, apart from the apex point $P_0$,
can be considered
as a 3-dimensional $(r , \theta, \phi)$-space, 
a Riemannian 3-space. 
It is in this 3-space, where the {\it integration}
in the {\it retarded potential} of our constraint equation 
will take place.

Three  of the components of this 3-space Riemannian metric
are given by the above construction,
\begin{eqnarray}
^{(3)}g_{r \theta} &=& 0, \quad \quad  ^{(3)}g_{r \phi} = 0,
\nonumber
\\
^{(3)}g_{rr} &=& 1.
\label{3.space.metric}
\end{eqnarray}
Proof of  the orthogonality 
$^{(3)}g_{r \theta} =  0$ and $ ^{(3)}g_{r \phi} = 0$: 
At fixed $r$, we have a $(\theta, \phi)$-2-sphere,
on which the radial distance from our observer at $P_0$ 
is independent 
of $(\theta, \phi)$. 
The orthogonality follows, 
because in a triangle with two equal sides $r$ from $P_0$ to
$P$ and $Q$ on the 2-sphere with an
infinitesimal basis $PQ$ (hence with an infinitesimal angle at the tip of
the triangle), 
the angles at the basis
of the triangle tend to $\pi/2.$ 

For   
formulating and solving the 
{\it angular constraint from  the past light-cone},
it is {\it irrelevant,} 
whether the {\it universe} is {\it approximately FRW}:
\begin{enumerate}
\item
In the angular momentum constraint, the input data of
the transvere components of $\vec{J}_{\varepsilon}$ 
on the past light-cone
are {\it averaged over all observation angles} using
    Eq.~(\ref{projection}).
Because of this angular averaging, 
it is {\it irrelevant,} whether the {\it universe} is {\it isotropic} 
around $P_0$ or not.
\item
In the angular momentum constraint from the past light-cone,
the input data $\vec{J}_{\varepsilon}$ 
must also be {\it averaged} over all {\it radial distances} 
on the past light-cone with the weight function discussed 
at the end of this paper.~---
Even a FRW universe is 
{\it not radially homogeneous on the past light-cone}, e.g.  
the universe at redshift $z = 15$ looks very different
from the universe at redshift $z = 0$.
\item
At every radial distance, 
the {\it observed radial velocities} of matter must be 
{\it averaged over angles}.
This gives an {\it expansion history} on the past light-cone.
For a parametrization of this expansion history, 
one will use the expansion history given by the fits to FRW models
given by WMAP 
and Planck. 
\end{enumerate}
\boldmath
\section{Inner geometry of past light-cone: 
unchanged by toroidal vorticity with $\ell = 1$}
\unboldmath

The precession of a gyroscope at $P_0$ can be caused
only by energy-current fields
in the toroidal vorticity sector with $\ell = 1$,
as shown in Sect.~\ref{gyroscope.precession.toroidal.vorticity}.

With the 
coordinatization-mapping $P \Rightarrow x^{\mu}_P$  
exactly fixed all over space-time
in the previous 
   section~\ref{sect.Mink.corridor},
the metric coefficients $g_{\mu \nu}$ 
are uniquely fixed by measurements.  

The energy-current field of matter, the input,
is in the toroidal vorticity sector
with $\ell = 1$. 
For linear perturbations, 
symmetry arguments will
force the geometric output $g_{\mu \nu}$
to be also in the same sector. 

On the past light-cone of $P_0$, we assume 
that the scalar sector is given by 
a spatially flat FRW background, and that the poloidal vorticity sector
and the tensor sector are zero.
Therefore the three components of the 4-metric, analogous to
Eq.~(\ref{3.space.metric}), are 
\begin{eqnarray}
^{(4)}g_{\chi \theta} &=& 0, \quad \quad  ^{(4)}g_{\chi \phi} = 0,
\nonumber
\\
^{(4)}g_{\chi \chi} &=& 0.
\label{4.space.metric}
\end{eqnarray}

Toroidal vorticity with $\ell = 1$ can only generate
{\it rigid rotations} on any 2-sphere. But rigid rotations leave
the components of the {\it metric on the 2-spheres unchanged}
from the FRW background (spatially flat) in
Eqs.~(\ref{metric.unperturbed.final}),
\begin{eqnarray}
&& {\mbox{including toroidal vorticity with}} \, \, \,  \ell = 1:
\nonumber
\\
&&g_{\theta \theta} = (a \chi)^2, \quad \, 
g_{\phi \phi} =   (a \chi \sin \theta)^2, \quad \,
g_{\theta \phi} = 0. 
\end{eqnarray}
Conclusions:
\begin{enumerate}
\item  
      For  toroidal vorticity  with 
               $\ell = 1$
      relative to a chosen observer,
      the {\it light-cones of this observer}
      have an {\it unperturbed inner geometry},
      and the coordinates can be chosen such that one has 
      {\it unperturbed metric coefficients} $g_{\mu \nu}$ 
      for the chosen light-cone coordinates
      $(\chi, \theta, \phi)$. 
\item 
      The {\it inner geometry} 
      of {\it all} light cones 
      (vertex at any space-time point) 
      remains
      {\it unperturbed} by 
           toroidal vorticity with 
               $\ell = 1.$ 
\end{enumerate}
%

\section{Evolution from light cone  to light cone} 
\boldmath
\subsection{The shift $\beta$}
\unboldmath

We now discuss 
the evolution from one light-cone to a neighbouring later light-cone 
(of our chosen observer) 
for the case of toroidal vorticity with $\ell = 1$.

The fundamental geometric quantity of this paper is the
shift $\beta$.

For a usual (3+1)-split with {\it space-like hypersurface-slices} 
labelled by a time
coordinate $t$, 
one defines the {\it shift} 3-vector-field and the {\it lapse} function
by considering the {\it connector}-4-vectors $\bar{C}_P$, 
where  (1) $\bar{C}_P$ is {\it normal} on the slice through $P$,
and (2)  
the 4-vector $(\bar{C}_P \, \delta t)$ {\it connects} slice
$\Sigma_t (P)$ with the slice $\Sigma_{t + \delta t}.$
See Misner, Thorne, and Wheeler
   \cite{MTW.shift}.

For unperturbed FRW with a {\it fixed-Hubble-time slicing} and 
with the {\it conformal-time} coordinate $\eta$, 
the connector 4-vectors are $\bar{C}(P) = \bar{e}_{\eta}(P)$.

At first sight, there {\it seems} to be a 
{\it problem} with  a (3+1)-split using light cones:
The {\it normals} $\bar{n}$ on  {\it light-cones}
are a multiple of $\bar{e}_{\chi},$ 
i.e. $\bar{n}$~lies in the tangent space to the light-cone and 
along a photon world-line.
Therefore  {\it normals on light cones do not connect}
successive light-cones.

However, there is {\it no problem for toroidal vorticity} perturbations.
In the {\it unperturbed} case,  the natural choice for the 
{\it connectors}
is 
$\bar{C} \equiv \bar{e}_v,$ 
where $\bar{e}_v(P) \delta v $ 
connects our observer's past light cones 
with $v = v (P)$ and  $v = v (P) + \delta v.$ 
This connector field $\bar{C} = \bar{e}_v$, which connects light cones,
is {\it identical} with the connector field $\bar{C} = \bar{e}_{\eta}$,
which connects fixed-Hubble-conformal-time slices.

The  {\it unperturbed lapse} function $N_0 \equiv \alpha_0$ is defined as the 
elapsed measured time  $\tau$ (proper time)  between
light-cones along the connector, i.e. for $(\chi, \theta, \phi) =$ fixed,
\begin{equation}
{\mbox{unperturbed lapse}} \, \equiv N_{(0)}(P) 
= (\partial \tau / \partial v)_P = a_P. 
\end{equation}

The {\it unperturbed shift} 3-vector $\vec{N}_{(0)}$
{\it vanishes}, because 
the unperturbed basis vector $\bar{e}_v(P) \equiv (\partial_v)_P,$
the tangent vector to the unperturbed $v${\it -coordinate line} 
is {\it not shifted away from}
the unperturbed {\it connector},  $\bar{C}(P) = \bar{e}_v(P),$
\begin{equation}
{\mbox{unperturbed shift}} \, \equiv \vec{N}_{(0)} = 0.
\end{equation}
The notation $N$ for the lapse function and
$N^i$ for the shift-3-vector
is from Misner, Thorne, and Wheeler,
    Ref.~\cite{MTW}. 

{\it Toroidal vorticity perturbations} are in the 3-vector sector,
therefore they
cannot produce a lapse perturbation $N_{(1)}$,
because the lapse function is a 3-{\it scalar},
\begin{equation}
 {\mbox{lapse perturbation}} \, \equiv N_{(1)} =  0.
\end{equation}

Because {\it toroidal vorticity fields}  
with $m = 0$ point in the $\phi${\it-direction},
the perturbed connector $\bar{C}$ can only aquire a $\phi$ component
(in addition to the unperturbed $v$-component),
and the {\it shift 3-vector} 
must point in the $\phi$-{\it direction}.
Therefore, 
all the action is in the  
$(v, \phi)${\it -tangent space}.
The connector 4-vector $\bar{C}(P)$  
and the shift-3-vector $\vec{N}(P)$ are defined by:
%
\begin{enumerate}
  \item  
     For infinitesimal $\delta v$,
     both  $\bar{C}(P) \, \delta v$ 
     and $\bar{e}_v(P) \, \delta v$  
     {\it connect} $P$ 
     with the neighboring coordinate line $v = v_P + \delta v$ in the 
     tangent space to the 
     $(v, \phi)${\it -coordinate surface},
\begin{equation}
C^v = 1.
\label{C.v.1}
\end{equation}
%
     Within this  (1+1)-dimensional $(v,\phi)$-tangent space,
     the {\it connector}  $\bar{C}_P$ is defined to be 
     {\it Lorentz-orthogonal}
     to  $\bar{e}_{\phi}(P)$, 
     \begin{equation} 
          g(\bar{C}, \bar{e}_{\phi}) = 0.
     \label{Lorentz.orthogonal} 
     \end{equation}
     %
  \item
     At each point $P$, the {\it shift}-3-vector $\vec{N}$
     is {\it defined} as the difference
     between   the 4-vectors
     $\bar{e}_{v}$ (tangent to the $v$-coordinate line)
     and the connector $\bar{C}$,
     \begin{equation}
     \vec{N} \equiv \bar{e}_{v} - \bar{C},
     \label{N.C}
     \end{equation}
     i.e.  $\bar{N}$ has no $v$-component from property (1),
and $\vec{N}$ is the shift of $\bar{e}_v$ 
     relative to the (1+1)-normal $\bar{C},$
where
we follow the {\it sign convention}  
of Misner, Thorne, and Wheeler \cite{MTW}.
  \item
     With $m = 0$ for toroidal vorticity,
     the shift-3-vector $\vec{N}$ can only have a $\phi$-component,
     which we denote by $N^{\phi} \equiv \beta$,
     \begin{eqnarray}
     m = 0 \, 
     &\Rightarrow& \, \,   {\mbox{shift}} \, 
     \equiv  \vec{N} = \beta \, \vec{e}_{\phi} \, \, \, 
     \Rightarrow \, \, \, N^{\phi} = \beta.
     \label{N.beta}
     \end{eqnarray}
     %

    A {\it positive shift}, $N^{\phi} > 0$, 
means that the {\it origin} of the $\phi${\it -coordinate}
is {\it shifted relative to the connector} 
in the {\it positive direction} 
with a {\it shift angle per unit conformal time}
$(d\phi/dv) = (d\phi/d\eta) = \beta$, hence
     \begin{eqnarray}
     C^{\phi }&=& \, (d\phi/dv)_{\rm (1+1)normal} \, = \, - \, N^{\phi}.
     \label{connector.phi.comp}
     \end{eqnarray}
\item For $\ell = 1$ and $m = 0$, the shift
is a {\it rigid rotation}
of the $(\theta, \phi)$-{\it coordinate system}  around the $z$-axis.
Hence, the shift function $\beta (v, \chi)$ 
is independent of $(\theta, \phi)$. The shift function is
the fundamental function for this paper. 
\end{enumerate}
%

The {\it lapse function} is defined as  
elapsed proper time~$\tau$ per unit coordinate time $v$ 
along the connector $\bar{C},$
%
\begin{eqnarray}
 {\mbox{lapse}}  \, \, 
&=&
(d \tau / dv)_{\rm (1+1)-normal} \,  
\equiv \, N \, = \, a. 
\end{eqnarray}

The angular velocity of a star measured by our observer at the origin 
is red-shifted from the value measured at the source.
But the {\it angular change per unit conformal time} 
measured by our observer at $P_0$ 
is equal to the value  
measured locally at the source $P$, 
\begin{eqnarray}
(d \phi / dv)_{{\rm obs.at}P_0} 
&=& (d \phi / dv)_{\rm locally \, meas. at \, source}. 
\nonumber 
\end{eqnarray}
%


\boldmath
\subsection{The perturbation of the metric and the inverse metric}
\unboldmath

From now on, 
we shall denote  unperturbed quantitites by~$(0)$ and
1st-order perturbations by $(1)$. 

For toroidal vorticity perturbations with $\ell =1,$
the perturbed metric components $g_{\mu \nu}$ 
must have one index~$v$, 
because perturbations only appear
in the {\it evolution} from one light-cone to a neighboring light-cone,
and one index~$\phi$, 
because  the {\it shift} is in the $\phi$-direction for $m=0$.
Hence, perturbations  
can only appear in $g_{v \phi}^{(1)}.$

The magnitude of the perturbation
$g_{v \phi}$
follows 
from:  
(1) the connector  $\bar{C}$
is  (1+1)-orthogonal to $\bar{e}_{\phi},$
  Eq.~(\ref{Lorentz.orthogonal}),
(2) $C^v =1$,
  Eq.~(\ref{C.v.1}),
and (3) 
$C^{\phi} =  - \beta$,
Eq.~(\ref{connector.phi.comp}),
\begin{eqnarray}
0 &=& g(\bar{e}_{\phi}, \bar{C}) 
= g_{\phi v}^{(1)} \, C^v + g_{\phi \phi}^{(0)} \, C^{\phi},
\nonumber
\\
     g_{v \phi}^{(1)} &=& \beta \, g_{\phi \phi}^{(0)} 
     = \beta \, \, a^2 \, \chi^2 \,  \sin^2 \theta.
     \label{metric.perturbed.FRW.final}
     \end{eqnarray}

The {\it line-element}
for toroidal-vorticity perturbations with $(\ell = 1, \, m = 0)$ 
on a  spatially flat FRW background 
in past-light-cone coordinates
follows from the last equation and from 
    Eq.~(\ref{metric.unperturbed.final}),
\begin{eqnarray}
ds^2   &=&  a^2(\eta)_{\eta=v-\chi} \, \,  [\, - dv^2 + 2 \, dv \, d\chi 
\nonumber
\\
&&  
+ \chi^2 (d \theta^2 + \sin^2 \theta \, d \phi^2)
    + 2 \, \beta  \, (\chi \sin \theta)^2 \, dv \, d\phi \, ],
\nonumber
\\
 \beta &=& \beta(v, \chi).
\label{perturbed.line.element}
\end{eqnarray}
%

To obtain the perturbation of the  inverse metric
one has to invert the (3x3)-matrix  
$g_{\mu \nu}$ in the $(v, \chi, \phi)$-sector,  
because the metric, 
   Eq.~(\ref{perturbed.line.element}), 
has no off-diagonal terms
involving $\theta.$ 
The inversion gives only one perturbed matrix element,
       \begin{eqnarray}
       g^{\chi \phi}_{(1)} &=& - \, \beta \, a^{-2},
       \end{eqnarray} 
where it is useful to remember:
       (1) upper indices $(\phi, \,  \chi)$, \,
       (2) minus sign, \, 
       (3) $a^{-2}$  multiplies $\beta$.


\boldmath
\section{Matter input measurable 
before solving Einstein's equations: $T^{v \phi}$}
\unboldmath
\label{input}

Einstein wrote in his letter to Felix Pirani of 2 February 1954 
as quoted by Ehlers 
in \cite{Barbour.Pfister.Ehlers}:
\begin{itemize}
\item ``If you have a tensor $T_{\mu \nu}$ and not a metric, then this does
  not meaningfully describe matter.
There is no theory of physics so far, which can describe matter without
already the metric as a ingrdient of the description of matter.
Therefore within existing theories the statement that the matter by itself
determines the metric is neither wrong nor false, but it is meaningless.''
\end{itemize}
From this argument, Einstein drew the conclusion that 
``one should no longer speak of Mach'sprinciple at all'',
quoted by Renn in \cite{Barbour.Pfister.Ehlers}.

Einstein's argument is utterly important, but it is {\it half-correct}
and {\it half-wrong}. It is our task to find out, 
{\it which component} of the energy-momentum tensor 
is 
(1)~measurable before having solved Einstein's
equations, hence before knowing the metric components, and
(2)~relevant for our problem.

Should we consider an {\it upper} or {\it lower} $\phi$-index
to have a {\it directly
measurable 3-momentum-input} on the matter side of Einstein's equations? 
Should we consider 
an {\it upper} or {\it lower} $v$-index 
to have a {\it directly measurable density}
on the {\it past light-cone}?

\boldmath
\subsection{The angular velocity index: upper $\phi$}
\unboldmath
For toroidal vorticity perturbations with $m=0$, 
the 3-velocity of matter is in the $\phi$-direction.
Should we consider an {\it upper} or {\it lower} $\phi$-index
to have a {\it directly
measurable 3-momentum-input} on the matter side of Einstein's equations? 
3-velocities and 4-velocities are prototypes for
the geometric object {\it vector} in the narrow sense,
for which the {\it natural} index-position  is an {\it upper index}.
A crucial observation was made for fixed $t$ in
Refs.~\cite{Rovaniemi, CS.1,CS.2}: 
The angular velocity 
$(d\phi/dt)$  can be {\it directly measured} (for nearby stars).
The 4-velocity is  
$u^{\phi} \approx d \phi/dt$ for non-relativistic motion of a star relative
to the unperturbed Hubble flow. 
Conclusion:
\begin{itemize}
\item 
   $u^{\phi}$ with the {\it upper index} $\phi$ 
   is {\it locally measurable input}, 
   the locally measured {\it angular velocity}
of a star around the $z$-axis. 
Conclusion: For $u^{\phi}$ with the {\it upper index},
Einstein's criticism is invalid. 
\item
   In contrast, $u_{\phi}$, with a {\it lower index}, 
   {\it cannot be used} as an {\it input} on
   the {\it matter-side} for {\it solving Einstein's equations},
because in  
$u^{(1)}_{\phi} = 
g^{(0)}_{\phi \phi} u^{\phi}_{(1)} + g^{(1)}_{\phi v} u^{v}_{(0)}$
the metric perturbation $g^{(1)}_{\phi v}$ 
{\it cannot be known} as an {\it input} 
without having already solved Einstein's equations all over the universe.
Conclusion: For $u_{\phi}$ with the {\it lower index},
Einstein's criticism is totally valid.
\item
   Conclusion: The index for the angular velocity (and angular momentum)
   around the $z$-axis must be
   an {\it upper} index $\phi$. 
\end{itemize}
Before our papers \cite{Rovaniemi, CS.1, CS.2}, 
{\it all} papers on Mach's principle missed the crucial
{\it super-Hubble-radius suppression}  of the dragging weight function,
because they used a lower momentum-index.

Note: 3-velocities and 4-velocities are prototypes for
the geometric object {\it vector} in the narrow sense,
for which the {\it natural, geometric} 
component-index-position  is an {\it upper index}.
{\it Without having solved Einstein's equations}, 
one does not know the metric, 
and one {\it cannot pull down an upper index}.

\subsection{The particle-density index: upper $v$}

The {\it particle number} (e.g. baryons or galaxies)  
on the past light-cone within  
a given $(\chi, \theta, \phi)$-{\it coordinate domain} 
is an observable, it is measured by astronomers.
It is given by  $n_{\alpha \beta \gamma}$, the
{\it particle-density 3-form} in $(\chi, \theta, \phi)$-coordinate space,
\begin{equation}
N_{\rm coord.domain} = \int_{\rm coord.domain}  n_{\chi \theta \phi} \,
d\chi \, d\theta \, d\phi.
\label{3.integral.over.3.form}
\end{equation}
Generally, a {\it 3-form} is defined in coordinate-components 
as a tensor with {\it three antisymmetric lower indices.}
An {\it integral} over a domain in three coordinates
calls for a 3-form in this coordinate basis 
as an integrand {\it without weight factors}, as shown in
   Eq.~(\ref{3.integral.over.3.form}). 
The {\it particle-density 3-form} $n_{\alpha \beta \gamma}$
gives 
the number of particles in the coordinate domain considered.

We use the 4-dimensional Levi-Civita tensor 
$\varepsilon_{\alpha \beta \gamma \delta},$
which is defined to be totally antisymmetric and to have
$\varepsilon_{0123} = 
\varepsilon_{v \chi \theta \phi} \equiv +\sqrt{- {\mbox{det}} (^4 g)} 
= a^4 \chi^2 \sin \theta$.
Using $\varepsilon_{v \chi \theta \phi}$, 
we can convert the particle-density 3-form 
to a particle-density contravariant vector-component  
$n^{v}$ with
$n_{\chi \theta \phi} \equiv n^v \, \varepsilon_{v \chi \theta \phi}  
= n^v \, (a^4 \chi^2 \sin \theta),$ hence
\begin{eqnarray}
&& N_{\rm coord.dom.}   
= \int_{\rm coord.dom.}    \, (an^v) \,
(a^3 \chi^2 \sin \theta) \,
d\chi \, d\theta \, d\phi,
\nonumber
\\
&&
\label{particle.number}
\end{eqnarray}
Conclusion:
\begin{itemize}
\item
The index for particle-density 
on the past light-cone must be an {\it upper} index $v$.
\end{itemize}
%


The {\it source-input} for Einstein's angular momentum constraint
which is {\it measured}
by our observer 
(with sufficiently precise apparatus in the future)
is the 
{\it angular change of position per unit conformal time}, 
$(d \phi / d v)$,
and the corresponding {\it energy-current component} $T^{v \phi}$ 
with {\it two upper indices},
which gives, for non-relativistic peculiar velocities,  
\begin{eqnarray}  
(\rho + p) \, (d \phi_{\rm matter} /dv) &=&  T^{v \phi} =  
 a  T^{t \phi} = a J^{\phi}_{\varepsilon}.
\nonumber
\end{eqnarray}
$T^{v \phi}$ can be measured 
{\it without prior knowledge} of the
{\it solution of Einstein's equations} all over space-time,
i.e. without prior knowledge of the metric field 
$g_{\mu \nu}$.
It follows that we must consider the Einstein equation 
$G^{v \phi} = 8 \pi G_{\rm N} \, T^{v \phi}.$~--- 
All other components 
of the energy-momentum tensor
$T^{\mu \nu}$ and
of the Einstein tensor $G^{\mu \nu}$ 
are unperturbed.

Because the matter-source of the relevant constraint equation 
is the angular momentum, this constraint equation is called the 
{\it angular momentum constraint}.

\section{Einstein's angular momentum constraint }

Apart from the FRW background,
for toroidal vorticity with $(\ell = 1, \, m = 0)$,
the Einstein-tensor has only the component $G^{v \phi}$, 
the energy-momentum has only the component $T^{v \phi}$,
and the  metric has only component $g_{v \phi}$. 

The Einstein equation 
for  $G^{v \phi}$ is,
\begin{eqnarray}
G^{v \phi} &=& 
8 \pi G_{\rm N} \, T^{v \phi}.
\nonumber
\end{eqnarray}

The computation of the exact $G^{v \phi}$ 
from the exact $g_{v \phi}$ for toroidal vorticity
with $(\ell = 1, \, m = 0)$
using  standard methods
will be documented in an appendix 
in an extended version of this paper.
The result with the sign convention 
of Misner, Thorne, and Wheeler \cite{MTW}
is,
\begin{eqnarray}
&&   a^4 \, G^{v \phi} = 
\nonumber
\\
&& =   \partial_{\chi}^2 \beta/2 
      + (\partial_{\chi}   \beta) (2 \chi^{-1} - {\cal{H}})
      +  \beta  (2 {\cal{H}}' - 2 {\cal{H}}^2).
\label{final.Einstein.tensor}
\end{eqnarray}
%

In the  energy-momentum tensor, 
$T^{\mu \nu} = (\rho + p) \, u^{\mu} u^{\nu}$, 
we consider 
{\it non-relativistic peculiar velocities} of
vorticity flows relative to the FRW background.
This implies that the perturbations are linear.~---
For toroidal vorticity with $(\ell = 1, \, m = 0)$, 
the energy-momentum tensor
has only one non-zero component
different from the FRW background (denoted by a subscript zero),
\begin{eqnarray}
a^2 \, T^{v \phi} &=& (\rho + p)_{(0)} \, (d \phi/dv)^{\rm matter}. 
\label{final.energy.momentum.tensor}
\end{eqnarray}

${\cal{H}}^2$ and ${\cal{H}}'$ are given in terms of $\rho$ and $p$ 
for a spatially flat FRW background, 
\begin{eqnarray}
{\cal{H}}^2 &=&  a^2 \, (8 \pi/3) \, G_{\rm N}  \, \rho,
\nonumber
\\
{\cal{H}}' &=&    
  -   a^2  (4 \pi / 3)  G_{\rm N}  (\rho + 3p),
\nonumber
\\
{\cal{H}}' - {\cal{H}}^2  
&=&  - \, a^2 \, 4 \pi G_{\rm N} \, (\rho + p).
\label{Hubble.and.Hubble.prime}
\end{eqnarray}

\section{Observer rotating 
relative to unperturbed FRW universe}

In this section, we specialize to an unperturbed FRW universe. 
There are no vorticity fields.

We no longer fix the orientation of the local spatial axes 
of the observer at $P_0$
to the observed directions to asymptotic quasars.
Instead, we fix the {\it orientation} of the {\it local spatial axes}
 of the {\it observer}
by {\it two local landmarks}.

For a consistency test,
we assume that the observer (with his local ortho-normal basis)
is {\it rotating relative} to the {\it FRW universe}
around his local $z$-axis with angular velocity,
\begin{eqnarray}
\Omega_{\rm observer \, rel. to \,  universe} &=&
- \, \Omega_{\rm matter \, rel. to \,  observer} 
\nonumber
\\
&=&  - \, a^{-1} \, (d \phi / dv)_{\rm matter}.
\nonumber
\end{eqnarray}

On the geometric side of Einstein's $G^{v \phi}$ equation, 
the shift function $\beta \equiv - \, (d \phi / dv)_{\rm (1+1)normal}$ 
is independent of~$\chi$, and the $G^{v \phi}$ equation reduces to
\begin{eqnarray}
G_{(1)}^{v \phi} &\Rightarrow& 
- \, 2 \, a^{-2} \, \beta \, (4 \pi G_{\rm N}) \, (\rho + p),
\nonumber 
\\
(8 \pi G_{\rm N}) \, T^{v \phi}_{(1)}
&=& a^{-2} \, (8 \pi G_{\rm N}) \, (\rho + p) \, (d \phi/dv)^{\rm matter},
\nonumber
\end{eqnarray}
hence, relative to the observer we have,
\begin{eqnarray}
\beta \equiv (d\phi/dv)_{(1+1)\rm normal} &=& (d \phi/dv)_{\rm matter}.
\label{exact.dragging.FRW}
\end{eqnarray}
This result proves {\it exact dragging of inertial axes} 
by matter in an {\it unperturbed  FRW universe}:
If all matter in the universe rotates rigidly
around the observer, then the {\it gyroscope axes}
(at the position of the observer)
are {\it exactly dragged} by the rotating matter.

This result is highly non-trivial: this result would not hold
for general relativity e.g. in a universe with the observed galaxies
out to redshift z = 1000 and no matter beyond.

As discussed in Sect.~IX of \cite{CS.1},
Einstein's equations together with given matter sources $T^{\mu \nu}$ 
are insufficient to obtain the geometry
in asymptotic Minkowski space.
General relativity (without explicit and totally 
non-trivial boundary conditions) 
is not invariant under going to a rotating coordinate system.

\section{Ordinary linear differential equation 
from vorticity with $\ell = 1$}

In this section, we consider the source-free,
{\it homogeneous} differential equation $G^{v \phi} = 0$ from
   Eq.~(\ref{final.Einstein.tensor}).
This source-free equation
will be needed for the
{\it Green function} away from the rotating spherical source-shell at
$\chi_{\rm source}.$

This equation can be solved numerically for any history of the 
Hubble rate.

For either a {\it cold-matter dominated} (CMD) universe
or a {\it radiation dominated} (RD) universe,
the homogeneous equation 
  Eq.~(\ref{final.Einstein.tensor})
becomes particularly simple,
\begin{eqnarray}
&& a (\eta)= \eta^P, 
\nonumber
\\
&&\quad \quad  {\mbox{observation event:}} 
\quad \quad \, \, \,  a_0   \equiv 1, 
\, \, \, \, \, \, \, \eta_0 \equiv 1,
\nonumber
\\
&& \quad \quad   {\mbox{big bang:}}
\quad \quad \quad \quad \quad \, \, \, \,  \, \, \,   \, a_{\rm BB} = 0,
\, \, \, \, \eta_{\rm BB} = 0,
\nonumber
\\
&& \quad \quad {\mbox{matter dominated:}} \quad \quad P = 2,
\nonumber
\\
&& \quad \quad {\mbox{radiation dominated:}} \quad P = 1,
\nonumber
\\
&& {\cal{H}}  =   P \, \eta^{-1},
   \quad \quad {\cal{H}}' = - P \, \eta^{-2},
\nonumber
\end{eqnarray}
On the light-cone of the observer, it is advantageous to use as the
independent variable 
the conformal time $\eta$ instead of the comoving distance $\chi$,
\begin{eqnarray}
&& {\mbox{light-cone of observation at}} \, \, P_0: 
\quad \, \, \,   \chi = 1 - \eta.
\nonumber
\end{eqnarray}
The homogeneous Einstein equation, $G^{v \phi} = 0,$ 
needed for the Green function away from the thin source-shell, becomes,
\begin{eqnarray}
&&  \partial_{\eta}^2\beta 
      -(\partial_{\eta}\beta)[4(1-\eta)^{-1}-2P\eta^{-1}]
      - 4\beta(P+P^2)\eta^{-2} = 0.
\nonumber
\\
&&
\end{eqnarray}
%

This ordinary linear differential equation of second order for the shift
$\beta(\eta)$ 
has 
{\it three singular points} of the
{\it regular type}, conventionally called ``regular singular points''.

The {\it definition} of a ``regular singular point'':
With a prefactor 1 for the second derivative $\beta'',$
the {\it prefactor} for the {\it first derivative} $\beta'$ has at most
a {\it single pole}, and the  
{\it prefactor} for the {\it function} $\beta$ has at most
a {\it double pole}, \cite{Abramowitz, Morse.Feshbach}. 

The important {\it result}: At {\it regular} singular points,
the solutions  have at most
{\it algebraic singularities}, $\beta(\eta) \propto \eta^{\alpha}.$

Any ordinary linear differential equation of second order 
with three regular singular points
is in the class of {\it Riemann's differential equation} 
    \cite{Morse.Feshbach, Abramowitz}.
Our differential equation for $\beta(\eta)$ 
has three regular singular points at
$\eta = 0$ (big bang), $\eta = 1$ (observation event), and
$\eta = \infty$ (infinite future).

The {\it exponents} of $\beta(\eta)$  for our differential equation are,
\begin{eqnarray}
{\mbox{big bang,}} &&  \eta  \rightarrow \, \, 0: 
\quad \quad \quad \quad \quad \beta \rightarrow \eta^{\alpha},  
\nonumber
\\ 
  (\alpha, \alpha') &=& 
 -P + 1/2  \pm   \sqrt{5P^2 + 3 P +1/4},
\nonumber
\\ 
{\mbox{obs. event,}} &&  (\eta - 1)   \rightarrow  \, \, 0: 
\quad \quad \beta  \rightarrow (\eta - 1)^{\bar{\beta}},
\nonumber
\\
(\bar{\beta}, \bar{\beta'}) &=& (0, \,  -3),
\nonumber
\\
{\mbox{inf. future,}} && (1/\eta) \rightarrow \, \, 0:
\quad \quad \quad \beta \rightarrow (1/\eta)^{\gamma}, 
\nonumber
\\ 
(\gamma, \gamma') &=& P + 3/2   \pm \sqrt{5 P^2  + 7P  + 9/4}.
\end{eqnarray}
We have
denoted Riemann's exponents at the observation point
by $(\bar{\beta}, \, \bar{\beta}')$ 
to distinguish them from our shift function $\beta(\eta).$ 
The sum of all six exponents must be equal to one
$\alpha + \alpha' + \bar{\beta} + \bar{\beta}' + \gamma +
\gamma' = 1$ for all Riemann differential equations.

At the observation event, 
the exponents are the same as in Minkowski space.
At the big bang, one exponent    
gives a {\it power-law suppression} in the
Green function 
for $\beta$, 
to be compared with the  exponential suppression
for large distances
found on a slice of fixed Hubble-time 
in~\cite{Rovaniemi, CS.1, CS.2}.

The regular solution at the observation event, $\eta = 1$,
goes to one, and is given by  hypergeometric series.
The solution decaying towards the big bang, after division by
$\beta^{\alpha}$, also goes to one, and it is  
given by another hypergeometric series
series.

Although we have not yet obtained the 
numerical solution for the Green function,
a cold-matter dominated
or a radiation-dominated universe,
nor for the more realisitic universe with dark energy plus cold dark matter
according to WMAP and Planck,
it is clear that the numerical solution exists and is well behaved.

From the Green function, one directly and simply 
obtains the dragging weight function
as in \cite{Rovaniemi, CS.1, CS.2},

From our analysis at fixed Hubble time \cite{Rovaniemi, CS.1, CS.2},
we expect that 
most of the dragging is done by matter around redshift $z \approx 1$.


\section{Proof of Mach's hypothesis 
for linear perturbations of FRW}

The {\it Green function} $G_{\beta} (\chi, \chi')$
for the shift $\beta$ is obtained 
by  solving Einstein's homogeous $G^{v \phi}$ equation,
as discussed in the last section.
This Green function, taken for $\chi$ infinitesimally close to the
observation event $P_0$ at $\chi = 0,$ 
gives directly the precession rate of
the gyroscope at $\chi = 0$ due to a rotating
source-shell at $\chi' = \chi_{\rm source}$ as discussed in \cite{CS.1, CS.2}.
In other words: this gives the {\it weight function for dragging}, $W(\chi),$
 obtained  for a fixed-Hubble-time slice in \cite{CS.1, CS.2}.

The weight function $W(\chi)$ 
for dragging by energy-currents on the past light-cone
depends on the history of the Hubble rate in the universe.
The dragging weight function is peaked near $z_{\rm source} =1$ relative to
the observation
\cite{Rovaniemi, CS.1, CS.2}.
For an observation time at today's redshift $\approx 20$, 
a good approximation
is a {\it matter dominated} universe. 
For an observation time at today's redshift $\approx 10'000$, 
a good approximation is a {\it radiation dominated} universe.
For observations today, a good approximation is a universe
{\it dominated by cold matter plus dark energy}.

The dragging weight function for these
three histories of the Hubble rate
will be evaluated numerically in a subsequent paper.
For a matter-dominated or a radiation-dominated universe, 
this involves nothing more than various hypergeometric series.

The resulting graphs will be instrucive.
But the {\it explicit forms} of the various dragging weight functions
are {\it not needed} for a {\it general proof} of 
{\it exact dragging} of inertial axes, i.e. the proof of 
the hypesis formulated by Mach.

The crucial observation: 
\begin{eqnarray}
&& \quad \quad \quad \quad {\mbox{from Eq.}}~\ref{exact.dragging.FRW} \, \, \, {\mbox{follows:}} 
\nonumber
\\
&&  {\mbox{dragging weight function}} \, \, W(\chi)) \, \,  {\mbox{normlized to unity,}} 
\nonumber
\\
&& 
\quad \quad \quad \int^{\rm big\, bang}_{\rm observation} d \chi \, W(\chi) = 1.
\label{weight.fct.normalized}
\end{eqnarray}
From this fact, Mach's principle follows directly,
\begin{eqnarray}
\vec{\Omega}_{\rm gyro} (P_0)
&=& 
 < \vec{\Omega}_{\rm matter}>_{W(\chi)}^{\rm past \, light \, cone}
\nonumber
\\
&=& \int_{\rm observation}^{\rm big \, bang} \, d \chi \, W(\chi) \, \, 
\vec{\Omega}_{\rm matter} (\chi).
\end{eqnarray}

Conclusions:
\begin{itemize}
\item  
The {\it hypothesis formulated by Ernst Mach} 
has been proved for all {\it linear} perturbations 
on the {\it past light-cone} of the observation
on spatially flat FRW backgrounds.
\item
The solution of the {\it angular momentum constraint} from the 
{\it past light-cone} gives {\it nothing more} and {\it nothing less}
than (1) the proof of {\it exact dragging} of 
{\it inertial axes} by {\it cosmic energy currents},
(2) the form of the {\it dragging weight-functions} 
for various Hubble-rate histories.
\end{itemize}
%






\bibliography{paperY.bib}

\end{document}